\documentclass{article}

\usepackage[final,nonatbib]{neurips_2018}

\usepackage[utf8]{inputenc} % allow utf-8 input
\usepackage[T1]{fontenc}    % use 8-bit T1 fonts
\usepackage{hyperref}       % hyperlinks
\usepackage{url}            % simple URL typesetting
\usepackage{booktabs}       % professional-quality tables
\usepackage{amsfonts}       % blackboard math symbols
\usepackage{nicefrac}       % compact symbols for 1/2, etc.
\usepackage{microtype}      % microtypography
\usepackage{graphicx}

\title{Studying oppressive cityscapes of Bangladesh}

\author{
  Halima Akhter \\
  KolpoKoushol\\
  Dhaka, Bangladesh\\
  \texttt{halimaakhter014@gmail.com} \\
  \And
  Nazmus Saquib\thanks{The author is also a member of the non-profit research lab KolpoKoushol. (Dhaka, Bangladesh).} \\
  MIT Media Lab\\
  Cambridge, MA 02139 \\
  \texttt{saquib@mit.edu} \\
  \And 
  Deeni Fatiha\\
  KolpoKoushol\\
  Dhaka, Bangladesh\\
  \texttt{dfatiha@mba2018.hbs.edu}\\
}

\begin{document}

\maketitle

\begin{abstract}
  In a densely populated city like Dhaka (Bangladesh), a growing number of high-rise buildings is an inevitable reality. However, they pose mental health risks for citizens in terms of detachment from natural light, sky view, greenery, and environmental landscapes. The housing economy and rent structure in different areas may or may not take account of such environmental factors. In this paper, we build a computer vision based pipeline to study factors like sky visibility, greenery in the sidewalks, and dominant colors present in streets from a pedestrian's perspective. We show that people in lower economy classes may suffer from lower sky visibility, whereas people in higher economy classes may suffer from lack of greenery in their environment, both of which could be possibly addressed by implementing rent restructuring schemes.
\end{abstract}

\section{Introduction}

\begin{quote}
    "I felt my lungs inflate with the onrush of scenery—air, mountains, trees, people. I thought, 'This is what it is to be happy.'" —Sylvia Plath, The Bell Jar
\end{quote}

The psychological influence of nature and environmental surroundings on the human mind is well-documented. Particularly, exposure to the sky and greenery seem to be of high importance to our psychological health and productivity \cite{asgarzadeh2009transdisciplinary}, \cite{chan2005communal}. Currently, there are many research projects to understand the effect of designing classrooms and work spaces with natural light and exposure to greenery. However, while wealthier schools and companies may have the resources to research and redesign interiors and exteriors to bring out the best in students and employees (for example, by designing rooftop and sky gardens), less affluent institutions cannot afford to do the same. This can potentially widen the gap in performance between more and less affluent schools and companies respectively.

It is important to remember, however, that wellbeing and productivity is not only a function of where people work but also where people live. In a densely populated city like Dhaka (Bangladesh), urbanism and scarcity of space created a surge of high-rise buildings in the recent decades. We demonstrate that in urban areas, higher income neighborhoods will have lesser amounts of greenery but access to more visibility to the sky, whereas lower income neighborhoods will have more greenery and lower visibility to the sky. We show this by developing a computer vision pipeline to use an existing 3D image database to extract and study sky visibility and the aesthetic nature of streets in Dhaka. To study the aesthetic nature of a cityscape, we focus our attention on the dominant colors present on sidewalks. 

Efforts to understand economies and aesthetic features of cities have been carried out before in the context of developed cities(e.g., \cite{doersch2012makes}, \cite{naik2014streetscore}). Many of these papers utilize computer vision techniques to extract features from crowdsourced building and city images, or use satellite images. Our work contributes some new findings in the context of a developing country. The implications of our pipeline and the study are multifold from the perspective of a developing country. Human behaviorial attitude associated with housing and real estate in the second densest city in the world, and the subsequent effect of urban planning and design on the housing economy can be studied using novel methods like computer vision. Traditionally, such studies have relied on surveys and census. We also show that metrics relating to natural and aesthetic beauty can be a possible factor in the housing economy of Dhaka. We discuss the implications of our results in section \ref{discussion}.

\section{Image collection and processing}
\label{gen_inst}

\subsection{Image collection}
\begin{figure}[h]
  \centering
  \fbox{\includegraphics[width=0.5\linewidth]{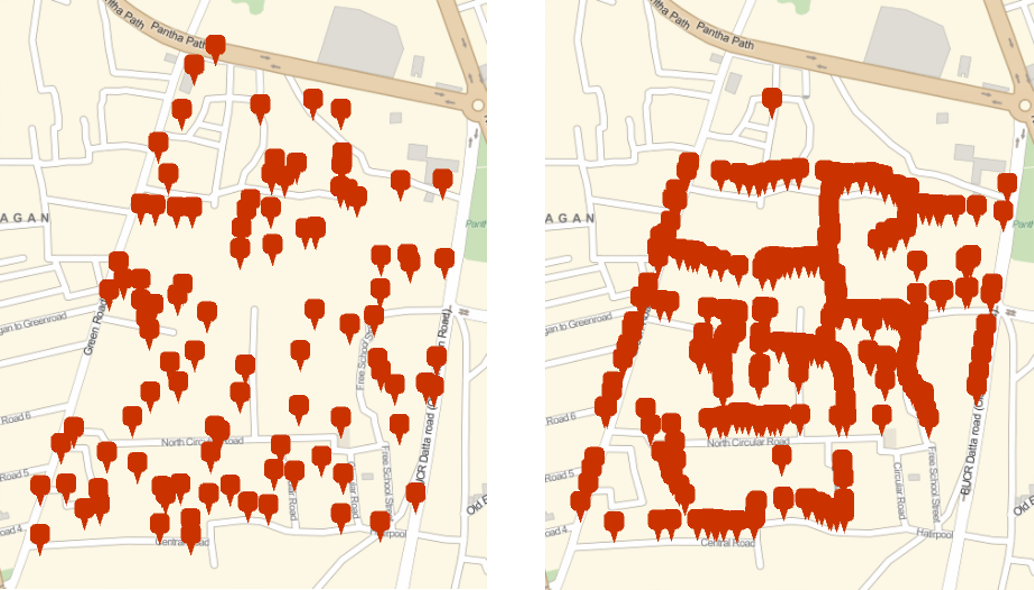}} 
  \caption{A set of sample points randomly generated within a ward polygon, and later snapped to the nearest roads.}
  \label{fig:map}
\end{figure}

Dhaka, the capital of Bangladesh, is divided into 92 geographical divisions known as "wards". We used Google Streetview API to collect the most recent (2016) images of the city. For each ward polygon, we sampled 300 points and snap them to the nearest road line using the Google Roads API. Figure \ref{fig:map} shows the process on a few blocks in a particular ward.

The Streetview API allows a user to point the camera direction at any angle. Assuming that the camera properties and image stitching process was consistent throughout the city, we pointed the camera angle to look upwards at the sky. We collected around 25k sky images in this way. For sidewalk images, we pointed the camera to look at the left and right side, and zoomed in to capture details such as banners, trees, store signs etc.

The Streetview image collection team did not enter some narrow streets where presumably their car could not fit. We ignore these streets in our analysis and focus on streets where cars and other forms of transportation can go through.

\subsection{Extracting sky}

\begin{figure}[h]
  \centering
  \fbox{\includegraphics[width=\linewidth]{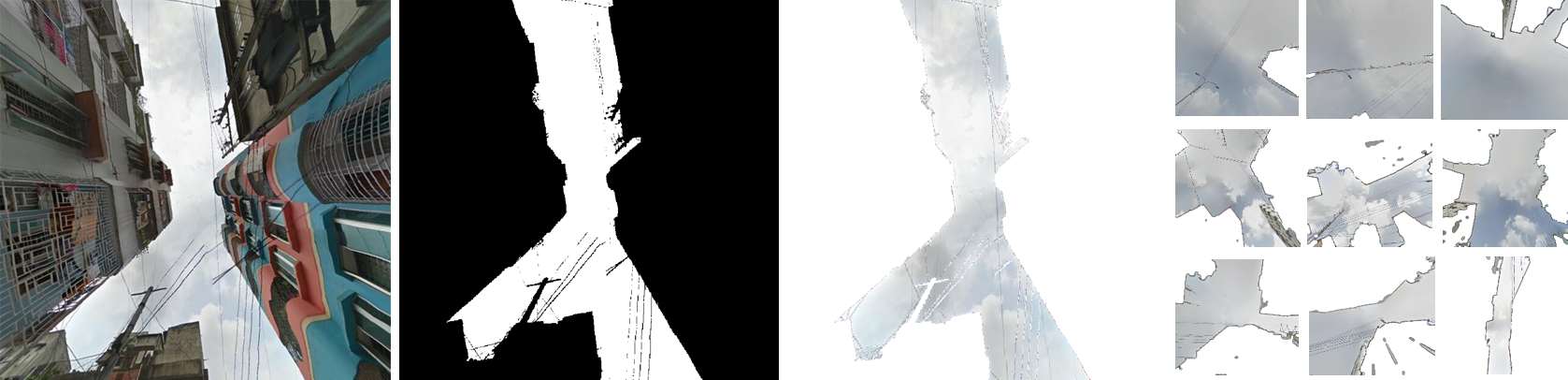}} 
  \caption{From left to right: (a) Streetview camera (pointed upwards towards the sky) produces image with sky and building borders, (b) segmenting sky from the rest of the image to create a mask, (c) using the mask to retrieve a sky footprint (or "skyprint"), and (d) a set of skyprints with varying sky visibility.}
  \label{fig:skyextract}
\end{figure}

We used an adaptive Gaussian Mixture model \cite{zivkovic2006efficient} to segment sky and buildings (or trees) lining the sky in an image. Figure \ref{fig:skyextract} shows the pipeline to extract a sky footprint (or "skyprint") from an image. A set of skyprints is included to show the different variabilities we noticed when mining the sky visibility from these images (figure \ref{fig:skyextract}(d)). There are narrow streets where the sky is barely visible due to high buildings, and also places where there are almost no objects in the skyline.

\subsection{Dominant color}
\begin{figure}[h]
  \centering
  \fbox{\includegraphics[width=\linewidth]{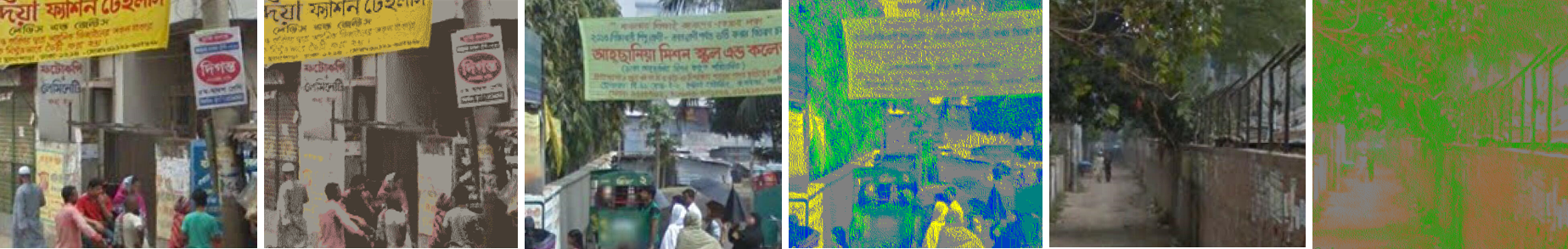}} 
  \caption{Dominant color extraction with (left) four colors, (middle) three colors, and (right) two colors.}
  \label{fig:domext}
\end{figure}

We used the k-medoid algorithm to cluster and find the dominant colors in the sideways image. Figure \ref{fig:domext} shows the result of using four, three, and two clusters on various images. After manual investigation of the results, we decided to extract three dominant colors for each sideways image we collected. For our subsequent analysis, we converted each image to HSB space from RGB space, and use the hue values.

\section{Sky view and rent economy}
\label{headings}

\begin{figure}
  \centering
  \fbox{\includegraphics[width=0.4\linewidth]{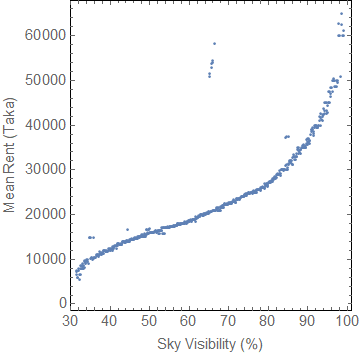}} 
  \caption{Mean rent vs. sky visibility scores.}
  \label{fig:rent_sky}
\end{figure}

To understand how sky visibility varies with rent, we generated a sky visibility score at each coordinate where the streetview image was acquired. Because the camera view properties were kept the same throughout, we received the same resolution image at each coordinate. The score was calculated as a percentage (the area of the image occupied by the sky with respect to the total area of the image). Next, we scraped rent prices from a local house rent website (pbazaar.com), finding house listings that are closest to the set of coordinates we used to collect the Streetview images, and assigned a rent value to each coordinate by interpolation. Then we tallied all the sky visibility scores and took the mean of the rent values at each tallied score. Figure \ref{fig:rent_sky} shows the mean rent value at each tallied sky visibility score.

We found a strong positive correlation between rent and sky visibility (0.859). Manual investigation of some of the outlier rent values revealed that these places were higher rent residential areas with moderate number of buildings. 

\section{Dominant color (greenery) and rent economy}
\label{others}

\begin{figure}[h]
  \centering
  \fbox{\includegraphics[width=\linewidth]{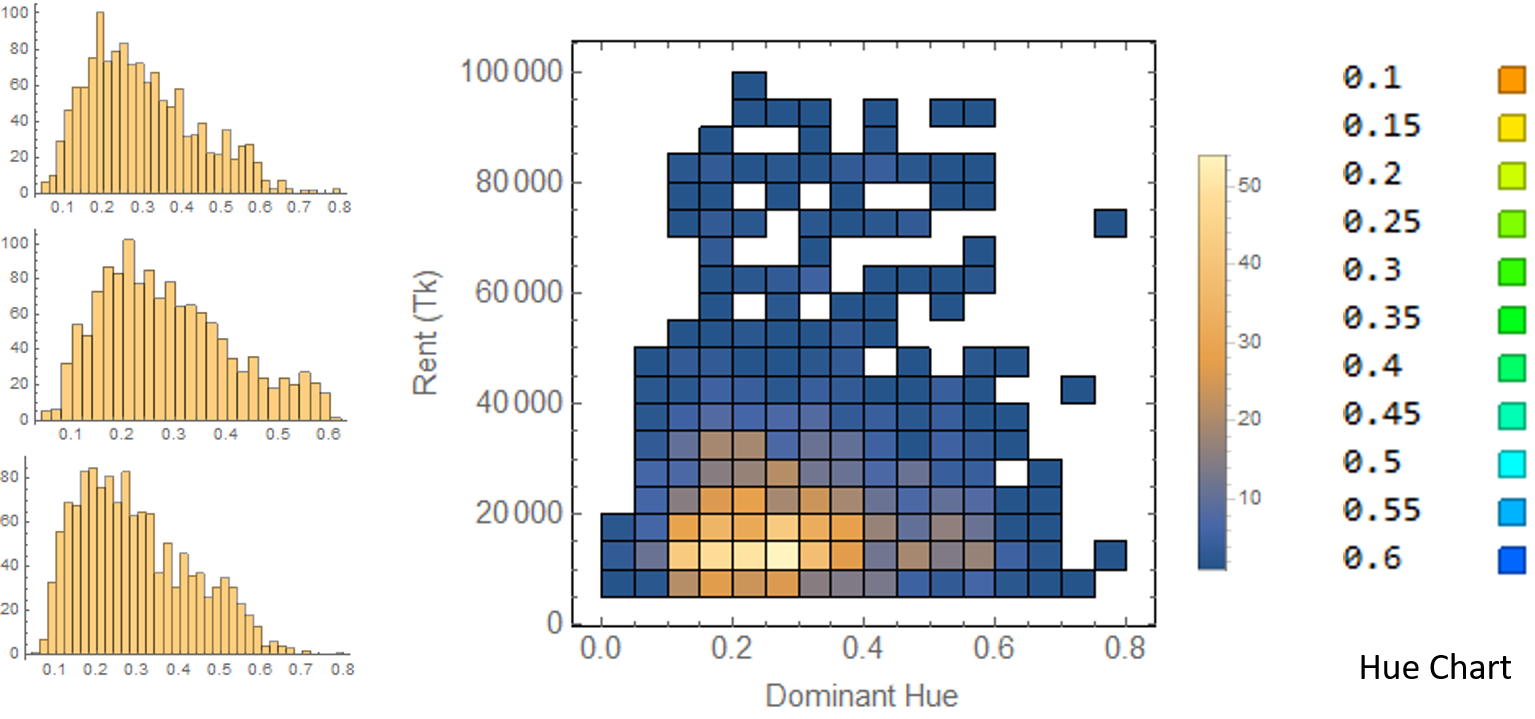}} 
  \caption{Dominant color statistics. (Left) histograms of the first three dominant colors's hue values found in all images. (Middle) a density histogram of the first dominant color's hue and mean rent in that area. (Right) a hue chart to interpret the hue values presented in the histograms. }
  \label{fig:huestats}
\end{figure}

As figure \ref{fig:huestats}(a) shows, the dominant colors mostly tend to fall in the green region. However, there is a good proportion of yellow and some red and blue hue too. We postulate that this is due to a large number of signs and banners on the street sidewalks, in many cases which are written using yellow and red paints (see figure \ref{fig:domext} for examples). Figure \ref{fig:huestats}(b) shows a density map of how the rent varies with the first dominant color. Most green areas enjoy a lower rent value, whereas higher rent areas tend to have less green colors in their street landscape. Interestingly, a good portion of areas where we find yellow and red color tones also tend to fall in the lower economy class.

\section{Discussion}
\label{discussion}

In general, such data and correlation require formal economics based investigation to come to any policy related conclusion, but we demonstrated that the ability to use common 3D street image databases to extract the skyline using computer vision techniques, and creating real estate data scraping pipelines will enable us to understand the wellbeing of citizens of all economic classes, and ask questions about the aesthetic values that we relate to, but are deprived of, in an oppressive urban landscape. Looking up at the sky and connecting with nature, a common innate need of human beings, seems to be a privilege in the new urban landscape, according to the primary analysis presented in this paper. This leaves the lower economy class with no choice but to detach themselves from viewing the sky. On the other hand, high economy classes tend to enjoy less greenery in their areas. Therefore this kind of technology can have interesting policy implications for increasing greenery where it is needed, and geographically precise rent restructuring interventions to include mental wellbeing in the housing economy. Closing such gaps could possibly mean a higher productivity, happiness, and lesser gap in performance among citizens.

\bibliographystyle{plain}
\bibliography{bib}
%[1] Doersch, C., Singh, S., Gupta, A., Sivic, J. and Efros, A., 2012. What makes paris look like paris?. ACM Transactions on Graphics, 31(4).

\end{document}